# Parallel Computing of Discrete Element Method on GPU[*]

Teruyoshi Washizawa, Yasuhiro Nakahara
*Simulation & Analysis R&D Center, Canon Inc., Tokyo, Japan*
washizawa.teruyoshi_at_canon.co.jp
nakahara.yasuhiro105_at_canon.co.jp

**Abstract**

General purpose computing on GPU for scientific computing has been rapidly growing in recent years. We investigate the applicability of GPU to discrete element method (DEM) often used in particle motion simulation. NVIDIA provides a sample code for this type of simulation, which obtained superior performance than CPU in computational time. A computational model of the contact force in NVIDIA's sample code is, however, too simple to use in practice. This paper modifies the NVIDIA's simple model by replacing it with the practical model. The computing speed of the practical model on GPU is compared with the simple one on GPU and with the practical one on CPU in numerical experiments. The result shows that the practical model on GPU obtains the computing speed 6 times faster than the practical one on CPU while 7 times slower than that of the simple one on GPU. The effects of the GPU architectures on the computing speed are analyzed.

## 1. Introduction

Recently, high performance computing on GPU in scientific and industrial fields has attracted much attention. Splendid achievements applied to various information processing problems, physical simulations, and social scientific problems can be found easily on the web site of NVIDIA [1]. The expected speed-up ratio reported in "Complete applications catalog" in [2] ranges from 1.5 to 500 depending on applications. As described later, since GPU obtains high computational performance by employing special hardware architectures and control processes, programmers are requested to tune their programs for fitting the specification of GPU. Fortunately, programmers can utilize NVIDIA's sample codes included in NVIDIA GPU Computing SDK obtained from the web site, which also demonstrate the superior performance of their products.

This paper focuses on discrete element method (DEM) which has properly used in particle motion simulation arising for the last decades [3, 4]. NVIDIA SDK provides a sample code for DEM and demonstrates the computing speed on GPU to be over 40 times faster than on CPU. However, the contact force employed in the sample code is too simple to use in practice. A practical DEM code has several kinds of force between a collided or interacted pair of particles. Hereafter, we refer to this model of force as practical model.

The first step to use the sample code in practical cases is to replace the NVIDIA's simple model with the practical model without modification of GPU-tuned parts. Numerical experiments are conducted to compare the computing speed of the following four cases, i.e., the practical model on CPU, the practical model on GPU, and the simple model coded in two languages on GPU. The result shows that the computing speed of the practical model is 6 times faster than that of the practical model on CPU while it is 7 times slower than that of the simple model on GPU. The effects of the GPU architectures on the improvement and the reduction in computing speed are analyzed.

The rest of this paper is organized as follows. The next section gives an overview of GPU architecture. Main features which affect the computational performance are described. The simple model and the practical model of the contact force in DEM are shown in section 3. Section 4 is devoted to the implementation of the practical model as well as the simple one. After describing the data structure, we show the process flow. Almost all processes are same for both models except the contact force calculation processes. The results of numerical experiments are shown in Section 5, which is followed by the discussion of the results.

## 2. An Overview of GPU Computing

---
[*]Special description of the title. (dispensable)

Before the detailed description of the implementation of the practical model, architecture of GPU is overviewed [5]. It helps us to analyze the result of numerical experiments described later.

In order to make the following description more comprehensive, a brief introduction of CPU architecture is given first. A CPU has several cores containing several, recently 6 or more processing units, an address counter and a decoder of instruction, a memory unit and other peripherals. Since one core has only one instruction address counter, processing units in each core are allowed to execute concurrently only one instruction. As described below, a GPU has 448 processing units which have their own instruction address counters so that they execute different sequences of instructions. This might be the biggest difference between CPU and GPU.

GPU also has hierarchical structure based on its processor hierarchy. GPU employs three types of hierarchy called thread hierarchy, memory hierarchy, and scheduling hierarchy. It employs a unique processing method called SIMT (single-instruction, Multiple-Thread) on these hierarchical structures.

Processor hierarchy is organized such that a GPU is composed of 14 streaming multiprocessors (SMs) consisting of 32 processing cores (PCs). GPU's SM and PC correspond to CPU's core and processing unit, respectively. The individual threads composing a warp start together at the same program address. They are free to branch and execute independently since they have their own instruction address counter and register state. Under the SIMT architecture, however, they are restricted to execute a common instruction at the same time because only one instruction interpreter is available per SM. This means that if they have M distinct instructions, these instructions are executed sequentially so that the speed up ratio is bounded to 32/M. This can be regarded as a kind of latency of instruction execution. This is called warp divergence and causes reduction of computational performance.

Thread hierarchy basically corresponds to the processor hierarchy, which is composed of thread, warp, and thread block. A thread is defined as the minimum unit of execution, which is equal to a sequence of instructions implemented by a user. A warp is defined as a set of 32 threads concurrently executed within a SM. A thread block is defined as a set of threads assigned to a SM. Since one SM executes one warp concurrently, the number of threads in a thread block is usually preferred to be a multiple of the number of threads in a warp. Thread blocks are organized into a data structure called grid. A GPU can execute two or more grids concurrently.

Memory hierarchy is composed of three types of memory, i.e., global memory, shared memory, and local memory. In the memory hierarchy, a global memory on a GPU is shared by all SMs while a shared memory on a SM is shared by all PCs. Moreover, each PC has its own local memory. From the global memory to the local memory, the access speed increases while the amount of memory size decreases. The shared memory can be accessed several 10 times faster than the global memory. Explicit description of usage of the shared memory enables us to enhance the efficiency of memory access. One more remarkable feature in memory access must be mentioned. When some threads in a warp are concurrently accessing addresses which are located locally in memory space, the memory access is improved. This function is called coalescing access.

The two-level distributed scheduler plays a great role to hide latencies caused by any instructions. At the chip level, a global work distribution scheduler assigns SMs to thread blocks. At the SM level, a dual warp scheduler in each SM distributes instructions in warps to its PCs. This hierarchical scheduling obtains highly effective load balancing among warps. However, it gives no contribution to reduce the warp divergence.

SFUs execute transcendental instructions such as sin, cosine, reciprocal, and square root. Each SFU executes one instruction per thread, per clock. All threads in a warp must share only 4 SFUs. If all threads must execute distinct special functions at the same time, only 4 threads are allowed to execute their special functions by using SFUs while other 28 threads must wait for the releases of the SFUs. This causes the speed up ratio of parallel computation is limited to the number of SFUs.

## 3. A Practical Model of Particle in DEM

NVIDIA's sample code employs a simple model of the force acting on a particle. The force acting on the j-th particle caused by the collision of the k-th particle, $F_S$ is defined as following:

$$F_S = k_{sp}\delta_n + k_{da}v + k_{sh}v_t + g \quad (1)$$

where $\delta_n$, $v$, $g$ is, respectively, displacement between two particles, velocity of a particle relative to the other one of a contact pair, and the gravitational acceleration. A suffix t and n denotes, respectively, the tangential and the normal component. The simple model is composed of spring force, damping force, sharing force and the gravity. All coefficients of RHS terms are constant. Although this model yields high computational performance of GPU, it is too simple to apply to practical simulations.

One of the standard model described in many articles related to DEM are much more complicated than the simple model in Equation (1) [6]. It is composed of a contact force $F$ and a contact torque $T$ described in the following equations,

$$F = F_t + F_n \quad (2)$$
$$T = r_i(n \times F_t) \quad (3)$$

where $F_t$ and $F_n$ are, respectively, the tangential and the normal component of the contact force defined as

$$F_t = -k_t \delta_t - \eta v_t, \quad F_n = -k_n \delta_n - \eta v_n \quad (4)$$

where $k$, $\eta$ and $r_i$ are the spring coefficient and the damping coefficient of particles and radius of the i-th particle, respectively. The tangential component of the contact force is adjusted if the following condition is satisfied,

$$F_t \leftarrow \mu |F_n|(F_t/|F_t|), \text{ if } |F_t| > \mu|F_n| \quad (5)$$

where $\mu$ is a coefficient of kinetic friction. The following equations give the definitions of the tangential component of velocity and the displacement between two particles, respectively:

$$v_t = v - (v \cdot n)n + (r_i \omega_i + r_j \omega_j) \times n, \quad (6)$$
$$\delta_t = \delta_{t,old} - (\delta_{t,old} \cdot n)n + v_t \Delta t, \quad (7)$$

where $\omega$, $\delta_{t,old}$, $n$, $\Delta t$ are angular velocity, displacement between two particles at the previous time, a unit vector between two particles, and a time step respectively. The coefficients of the terms in RHS of Equation (4) is, respectively, defined as functions of the physical properties,

$$k_t := C_{k,t}(i,j)\sqrt{|\delta_n|/(r_i^{-1} + r_j^{-1})}, \quad (8)$$
$$k_n := C_{k,n}(i,j)\sqrt{|\delta_n|/(r_i^{-1} + r_j^{-1})}, \quad (9)$$
$$\eta := \alpha(i,j)\sqrt{k_n/(m_i^{-1} + m_j^{-1})}, \quad (10)$$

where $C_k(i,j)$, $\alpha(i,j)$, $m_j$ are spring parameter, restitution parameter, and mass of the j-th particle, respectively. Hereafter, we refer to all variables in above equations as particle properties.

## 4. Implementation

NVIDIA's sample code has been tuned so well as to demonstrate the speed up ratio of over 40 than an ordinary CPU. This performance must be maintained through implementation of practical DEM code. The first step we decided to take is to replace the simple model in the sample code with the practical model in Equations (2) to (10). This implementation can be used as a reference of the lower bound of the DEM performance for the practical model on GPU. When we apply some approximations to speed up, we can evaluate the ratio of both the computing speed and the approximation error to those of the lower bound case.

Another factor we have to determine in advance is a programming language. NVIDIA provides a programming environment named CUDA (Compute Unified Device Architecture) [5] which is only available for NVIDIA GPUs. On the other hand, OpenCL has been proposed to platform-independent programming for GPUs. We choose OpenCL to implement the practical model because of its performance-portability [7]. NVIDIA provides sample codes for both languages.

The practical model is also implemented in C++ for execution on CPU. Since the contact force is often arranged in practical applications, the comprehensive interface to add or delete a force term should be implemented. Object-oriented programming enables us to obtain such maintainability. The C++ implementation is a reference as the most maintainable case.

### 4.1. Allocation of Variables

To detect particle collisions effectively, DEM simulation usually uses a collision detection grid (CDG). The CDG divides the whole computational space into small sub-regions called cell. Every particle is registered to a cell that occupies its location. It is guaranteed that all particles registered in a cell are close each other. Particles colliding with the j-th particle at the next time step can be found in either the cell containing the j-th particle or its neighbor cells. For example, the number of cells to be searched is $3^3$ in 3-dimensional space. The candidates of colliding pairs decrease as the number of cells of the CDG increases because the number of particles in each cell reduces on average. Hence, the CDG much reduces the computational time to search colliding pairs. It should be noticed, however, that, if the size of the buffer of particles in a cell is greater than the cache memory size, the computational performance is considerably degraded. The number of cells should be selected by considering these factors. The sample code employs the CDG structure.

Most straightforward way to manage particles in the CDG is to assign an array to each cell to register particles referred as "array-in-grid" [3]. Since this method is obviously memory consuming, improved methods have been proposed [3]. The sample code uses a correspondence map from particle to cell. The corresponding map CM is a 1D array set to be CM[j] = k when the j-th particle is inside the k-th cell. The number of entries of CM is equal to the number of particles.

Locations of variables in memory space affect the computational performance. Variables located consequently in memory space can be uploaded once on high-speed accessed cache memory. If all variables needed for calculation of some value are uploaded on the cache memory at the same time, the minimum memory access time is needed. Otherwise, time-consuming memory access reduces the computational performance.

The sample code has the process of sorting the particles along the correspondence map. By performing this process, a pair of particles close to each other in calculation space is also located to be close in memory space. It follows that a collision pair of particles is expected to be close in memory space so that they are uploaded simultaneously on cache memory when the contact force is calculated. The sample code uses the bitonic sort algorithm which could be the best for GPU.

### 4.2. Process Flow

Both the NVIDIA's simple model and the practical complicated one are obtained by executing along the same flow of operations except the calculations described in Equations (1) – (10). The process flow is shown below [8]:
1. Update the all particle properties.
2. Make the correspondence map from a particle to a cell. The correspondence map is contained in an integer array CM. CM[j] = k when the j-th particle is inside the k-th cell.
3. Sort the CM to get the sorted CM, SCM and the map from SCM to CM, SCCM. These arrays satisfy the following relationship:
$$SCM[j] = CM[SCCM[j]]. \qquad (11)$$
4. Reorder all the properties along SCM by bitonic sort.
5. Assign the i-th thread to the SCM[i]-th particle.
6. Collect the SCM[i]-th particle's cell and its neighbor cells. If the particle's cell is the (i,j,k)-th cell in 3D computational space, a set of these cells is
$$\{(l,m,n) | i-1 \leq l \leq i+1, j-1 \leq m \leq j+1, k-1 \leq n \leq k+1\}. \qquad (12)$$
7. Calculate the contact force between two particles in Equation (1) for the simple model, or Equations (2) to (10) for the practical model.
8. Calculate the contact force between particles and walls for the practical model. A wall can be modeled as a particle with the infinite value of radius.

The difference between the simple model and the practical model is only the calculation of force in Step-7 and Step-8. Before the calculation of contact force in Step-7, collision detection is executed for any pair belonging to the cells collected in step-6. The same detection is performed for any pair of a particle and a wall in Step-8. This detection generates huge amount of load unbalance causing the reduction of computational performance.

The practical model is implemented in OpenCL and C++ for GPU and CPU, respectively. The implementation in C++ on CPU is a kind of reference of the best code for maintainability for modifying the contact force composition.

### 5. Numerical Experiments

Numerical experiments were conducted to compare the performances of the following four cases, the simple model coded in CUDA on GPU, the simple model coded in OpenCL on GPU, the practical model coded in OpenCL on GPU, and the practical model coded in C++ on CPU. First two cases have been provided by NVIDIA as sample codes. The remains are coded by the authors in OpenCL and C++, respectively. The first two cases are used to find the effect of the difference of programming language for the simple model. The second and third cases show the effect of the difference of the contact force model. The difference of the performance between GPU and CPU for the practical model is evaluated from the comparison of the last two cases. The first and the last case are, especially, regarded as references of the fastest and the most maintainable model, respectively.

A sample problem simulates a time evolution of $2^{17}$=131,072 particles with same radius which fall from a box on a floor through a slit at the bottom of the box. The time evolution is continued until displacements of all particles are less than a predetermined value. Specifications of CPU and GPU are shown in **Table 1** and **Table 2**, respectively. Computing speed [particles/sec] of these four cases are shown in **Table 3**.

The difference between first two cases from the top in **Table 3** is only the programming languages used and has little effect on the computing speed.

The effect of the difference of the contact force model on computing speed can be evaluated by comparing the second and third cases. The computing speed of the practical model on GPU is 7 times slower than the simple model on GPU. The effect of the difference of processor type and of programming language on computing speed can be also evaluated by comparing the third and fourth cases. The computing speed of the practical model on GPU is 6 times faster than the practical model on CPU. This concludes that the GPU computation is available in practical use even if a complicated practical model of contact force is employed.

### 6. Discussion

Here we consider factors which effect on two cases, one is when the practical model on GPU has greater performance than that on CPU and the other is when the practical model on GPU is less performance than the simple model on GPU. Effects of factors to the computing speed are estimated under the assumption that all factors are independent to each other. The integrated effects of all factors evaluated as the product of all estimates under this assumption give the bounds for both cases.

First, let us focus on the improvement of the computing speed by replacing CPU with GPU. See the first and the second rows from the bottom in **Table 3**. As described above, we can find that the practical model on GPU is 6 times faster than that on CPU. The following three factors mainly affect the improvement.

First, a GPU has 448 PCs (14 SMs per GPU times 32 PCs per SM) which execute concurrently. If all threads execute instructions independently, the computing speed is expected to be increased by 448/(2.93/1.15)/4=44 (the number of PCs divided by the clock frequency ratio and by the number of processing units in one CPU-core). This astonishing improvement is the best case in which all pairs containing the neighbor cells in Equation (12) have collisions. This case, however, can be seldom seen in practice. As the opposite limit, consider case when, among all threads in a grid, only one thread needs to calculate the contact force while the particles assigned to the other threads have no collisions. This is obviously the worst case when the computing speed is same as CPU. All practical cases are located between these two cases. It can be said that the probabilities of collisions of any pairs increase as the density of particles increases. In the application used for our benchmark, particles are located so densely that most particles are expected to have collisions with their closest particles. The effect of this kind of warp divergence can be evaluated. A thread calculates the contact force acting on a particle after detecting the collision with other particles in the collection of 27 cells described in Equation (12). Warp divergence in this case is generated by the difference of the number of particles being checked and by the necessity of contact force calculation for each pair. If all particles have the same diameter $d$, the number of particles closest packed into a cell is given as

$$\frac{L_x}{d}\frac{L_y}{\sqrt{3}d/2}\frac{L_x}{\left(\sqrt{2/3}d\right)} = \frac{\sqrt{2}L_xL_yL_z}{d^3} \quad (13)$$

where $L_x$, $L_y$, and $L_z$ are the length of edges of a cell along x-, y-, and z-axis, respectively. In the NVIDIA's sample code, all lengths of edges are set to be equal to the diameter of a particle. This reduces Equation (13) to $\sqrt{2}$. The maximum number of particles being checked is, therefore, 27 times $\sqrt{2}$, or about 47. On the other hand, the maximum number of particles collided with one particle simultaneously can be evaluated to be 12. Hence, in case of densely distributed particles, a particle is collided with about a quarter of all particles contained in a cell. This means a thread is idle for three quarters of whole computing time. Since the collided pairs of each particle probably be different from others, all threads must perform the contact force calculation for all pairs regardless of their validity under the SIMT architecture.

**Table 1. specification of CPU.**

|  | Specification |
|---|---|
| Model name | Intel®Xeon®CPU X5670 |
| Clock frequency | 2.93GHz |
| Cache | CPU: L1 I cache, 32K, L1 D cache; 32K<br>CPU: L2 cache, 256K<br>CPU L3 cache, 12288K |

**Table 2. specification of GPU.**

|  | Specification |
|---|---|
| Model name | C2050 (Fermi) |
| Number of multiprocessors | 14 |
| Total number of threads | 448 |
| Clock frequency of processing cores | 1.15GHz |
| Maximum band width | 144 GB/s |
| Peak performance of Double Precision(FMA) | 515.2 GFLOPs |

**Table 3. Comparison of the computing speed of four cases.**

| Computational model | Programming language | Processor type | Computing speed $10^3$[particles/sec] |
|---|---|---|---|
| NVIDIA's simple model | CUDA | GPU | 22,359 |
|  | OpenCL | GPU | 21,266 |
| Practical model | OpenCL | GPU | 2,960 |
|  | C++ | CPU | 474 |

The warp divergence in this case makes the computing speed 4 times slower. Notice that the reduction rate would be up to 47 when only one pair has a collision. It should be also noticed that the total number of particles in a cell effects on warp divergence strongly. When the computing time of the contact force calculation is much larger than that of the CDG registration, the cell size in the NVIDIA's sample code must be the best choice. This is because all the collided particles should be found in the neighbor cells and the ratio of the collided particles to the total ones in a cell is expected to be large.

Second, a large amount of threads increases the effective memory access speed. The two-level distributed scheduler arranges uncompleted threads so that the total amount of idol time of all threads would be minimized. If this scheduling works perfect, the latencies caused by the conflicts of memory access and of SFU operations are completely hidden.

Third, coalescing access increases the memory access speed. In our implementation, all particle properties are reordered according to the locations of particles. Variables used for calculating the contact forces are, therefore, located locally in memory space. This enables coalescing memory access.

Next, the reduction of computing speed by replacing the simple model with the practical one is considered. See the second and the third rows from the top in **Table 3**. We can find that the practical model on GPU is 7 times slower than the simple model on GPU. The major causes of this reduction are the following three factors.

First, increment of the amount of computation of the contact force proportionally increases the computing time. Second, increment of the number of if-branches increases the warp divergence. N-layered nested branches generate $2^N$ paths. Since the different paths must be executed sequentially, the computational time increases proportionally to the number of paths. Third, usage of the SFUs causes the pause of operations. If all 32 threads in a warp must calculate special functions at the same time, requests from only 4 threads are accepted and the remaining 28 threads must wait for the releases of SFUs. In this case, the computing speed reduces by at most 8 times than those without using the SFUs. The practical model indeed includes square root functions in Equations (8) to (10), which request to use the SFU twice every calculation of the contact force of a particle.

## 7. Conclusions

We reported the computational performance of GPU for practical DEM computation by comparison with CPU. Since the model of contact force in NVIDIA's sample code is too simple to evaluate the practical performance, the simple model was replaced with a practical model used in many DEM simulations. The computing speed of the practical model on GPU obtains 6 times faster than that on single core CPU while 7 times slower than the NVIDIA's simple model on GPU. Because of no approximation and no special tricks in our code, this computing speed is regarded as the lower bound of DEM simulation on GPU. The effects of the GPU architectures on the computing speed were analyzed for the further improvement. Especially, branch divergence among threads in a warp caused by the branches in contact force calculation was discussed.